\documentclass[a4paper,11pt]{article}

	\usepackage{enumerate,hyperref,siunitx}
	\usepackage{enumitem,}
	\usepackage{siunitx}
	\usepackage{cite}
	\usepackage{bm,amsmath,amssymb,physics}
	\usepackage{graphicx}
	\usepackage[font=small,labelfont=bf,figurename=Figure,labelsep=period]{caption}
	\usepackage{multirow,tabularx}
	\usepackage[margin=1in]{geometry}
	\usepackage{float,pdflscape,authblk,setspace,lineno}	
	\usepackage{xcolor}
	\usepackage{amsthm}

	\usepackage{cleveref}

	\newcommand{\Xt}{\mathbf{X}(t)}
	\newcommand{\Wt}{\mathbf{W}(t)}
	\newcommand{\Xz}{\mathbf{X}_0}

\begin{document}

	\title{Smoothing in linear multicompartment biological processes subject to stochastic input}


	\author[1]{Alexander P Browning}
 	\author[2]{Adrianne L Jenner}
 	\author[1]{Ruth E Baker}
 	\author[1]{Philip K Maini}

	\affil[1]{Mathematical Institute, University of Oxford, Oxford, United Kingdom}
 	\affil[2]{School of Mathematical Sciences, Queensland University of Technology, Brisbane, Australia}
	

\date{\today}
\maketitle
\footnotetext[1]{Corresponding author: browning@maths.ox.ac.uk}

\vfill
	\renewcommand{\abstractname}{Abstract}
	\begin{abstract}
		\noindent
		Many physical and biological systems rely on the progression of material through multiple independent stages. In viral replication, for example, virions enter a cell to undergo a complex process comprising several disparate stages before the eventual accumulation and release of replicated virions. While such systems may have some control over the internal dynamics that make up this progression, a challenge for many is to regulate behaviour under what are often highly variable external environments acting as system inputs. In this work, we study a simple analogue of this problem through a linear multicompartment model subject to a stochastic input in the form of a mean-reverting Ornstein-Uhlenbeck process, a type of Gaussian process. By expressing the system as a multidimensional Gaussian process, we derive several closed-form analytical results relating to the covariances and autocorrelations of the system, quantifying the smoothing effect discrete compartments afford multicompartment systems. Semi-analytical results demonstrate that feedback and feedforward loops can enhance system robustness, and simulation results probe the intractable problem of the first passage time distribution, which has specific relevance to eventual cell lysis in the viral replication cycle. Finally, we demonstrate that the smoothing seen in the process is a consequence of the discreteness of the system, and does not manifest in system with continuous transport. While we make progress through analysis of a simple linear problem, many of our insights are applicable more generally, and our work enables future analysis into multicompartment processes subject to stochastic inputs.
	\end{abstract}
	\vfill

	\renewcommand{\abstractname}{Keywords}
	\begin{abstract}
		\noindent 
		\centering
		multicompartment, Ornstein-Uhlenbeck, smoothing, robustness, cell lysis, virus life cycle
	\end{abstract}
	\vspace{1cm}
	\vfill

\clearpage
\section{Introduction}

\begin{figure}[!b]
		\centering
		\includegraphics[width=\textwidth]{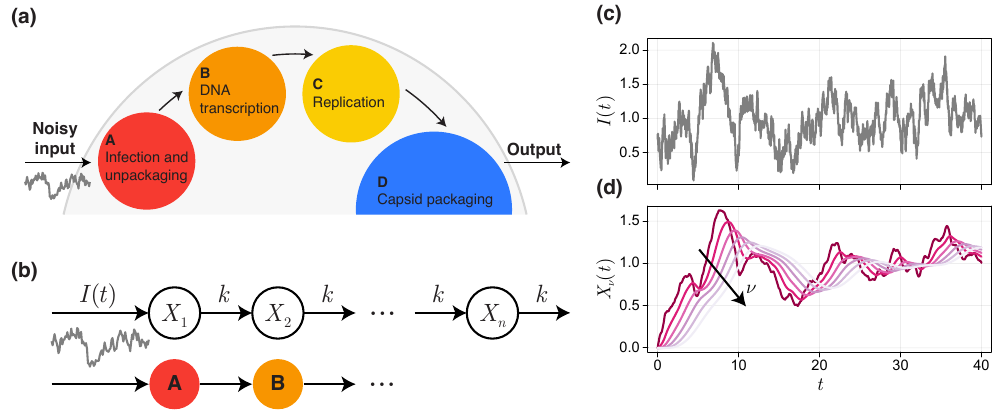}
		\caption[Figure 1]{{\bfseries Multicompartment model of viral replication subject to stochastic input.} (a) The general viral replication cycle. Virions enter a cell and progress through a multistage process, before accumulating  following repackaging. (b) We study a linear analogue of the virus problem, namely a system subject to random input, $I(t)$, modelled as an Ornstein-Uhlenbeck process with mean $\mu$, reversion strength $\theta$, and noise magnitude $\sigma$. A realisation of the input is shown in (c). Material then progresses through $n$ compartments at constant rate $k$. (d) A realisation of the system in (b), where $X_\nu(t)$ models the concentration of material in each compartment $\nu = 1,2,\dots,6$. It is evident that passage through the compartments has a smoothing effect.}
		\label{fig1}
	\end{figure}

Many biological processes comprise multiple independent stages. Viral replication, for example, is a multistage process whereby virions enter a cell through endocytosis, are unpackaged before DNA replication, repackaging, and release (\cref{fig1}a) \cite{Cann.2008,Louten.2016,Sazonov.2022}. Similar multistage processes are evident in bacteriophage replication \cite{Campbell.2003} and progression through the cell cycle \cite{Morgan.2007}, pervasive at the molecular (i.e., cascade reactions \cite{Huang:1996}) and macroscopic (i.e., transport through discrete layers \cite{Carr.2019df}) levels, and even manifest in social processes such as queuing \cite{Liu.2004}. A challenge for many systems is to modulate the impact of what are often highly variable external environments. For instance, while the intermediate stages of viral replication may be optimised to achieve high levels of virion multiplication, the system has either no, or only very limited, control over the number of virions entering the cell \cite{Schulte.2014,Jones.2021}. For lytic viruses, should the number of virions present inside a cell exceed capacity the cell will lyse, destroying the system and ceasing replication \cite{Heaton.2017}.

The time until cell lysis---more broadly, the time until the first occurrence of any event within a stochastic process---can be modelled as a \textit{first passage time} (FPT) and is dependent, among many other factors, on the variability and the autocorrelation of the process \cite{Redner.2001,Kannoly.2022df}. Statistics such as the variance, autocorrelation function, and the FPT, are commonly studied in scalar stochastic systems in biology \cite{Redner.2001,Allen:2011}. Many linear and non-linear systems described by continuous or discrete-space random walks have closed form solutions available for the aforementioned statistics, and if not for the FPT distribution itself, then for the mean, variance, and higher order moments of the FPT \cite{Redner.2001,Simpson.2015,Rijal.2022}. Analysis of higher-dimensional systems (i.e., described by more than one first-order differential equation) has, to date, been restricted to a fixed number of dimensions; most commonly two or three, where the velocity or acceleration of a particle is described by a stochastic process \cite{Bernard.1972,Grigoriu.2020}. General techniques for analysis, such as through the Fokker-Planck equation, quickly suffer from the curse of dimensionality, and dimension reduction techniques may yield lower-dimensional stochastic processes that lack the Markov property typically exploited in analysis.

Motivated in particular by the viral replication cycle, in this work we study the properties of a linear $n$-compartment model subject to an independent external input, which we model using a mean-reverting Ornstein-Uhlenbeck process (\cref{fig1}b). While the presentation and analysis of stochastic models is ubiquitous throughout the biological literature, efforts have been largely restricted to the study of the effects of intrinsic noise, process noise, and fluctuating parameter values, including for analysis of viral replication dynamics \cite{Matis.1979,Yu.2004,Donnet.2013,Mistry.2018,Teufel.2020}. Meanwhile, there has only been a very small amount of work focussed on analysis of the statistical properties of largely deterministic systems subject to independent stochastic input \cite{Anderson.2007}. In the context of control and state-space estimation, our model parallels highly studied linear state-space models such as autoregressive models and the linear Kalman filter \cite{Box.2015}. Given the scarcity of general knowledge related to the behaviour of multicompartment models subject to input noise, we are motivated to study a simple linear analogue of the viral multicompartment problem subject to a single Ornstein-Uhlenbeck input.

The choice to study a simplified linear stochastic model allows us to formulate the multicompartment problem as a multidimensional Gaussian process, enabling us to draw on the significant body of literature devoted to study of the statistical properties of such systems \cite{Meucci.2009,Vatiwutipong.2019,Gardiner} to formulate a series of analytical expressions for key statistics unique to stochastic processes including the variance, covariance, and autocorrelation function. Alternative approaches to a more theoretical analysis could include the study of system response to pure waves and input pulses; however, the goal of this work is to study the simple model directly. While we are not able to solve explicitly for the probability density function of the FPT, we present a series of numerical and approximate results that provide insight into the FPT, and the rate at which the mean FPT scales with the number of compartments in the system. We then apply our linear model to study how the behaviour or robustness biological systems can be modified through perturbations to unidirectional progression through the system. Viral replication, for example, is known to be a highly stochastic process, and progression through replication stages is very often not unidirectional \cite{Louten.2016}.

\section{Mathematical model}

The problem presented in \cref{fig1}b can be expressed as the linear system of stochastic and ordinary differential equations
	\begin{equation}\label{sys}
	\begin{aligned}
		\dd I(t) 	 &= -\theta\Big(I(t) - \mu\Big) \,\dd t + \sigma \,\dd W(t),\\
		\dd X_1(t)   &= \Big(I(t) - kX_1(t)\Big) \, \dd t,\\
		\dd X_\nu(t) &= \Big(k X_{\nu-1}(t) - k X_\nu(t)\Big)\,\dd t,\qquad \nu = 2,\dots,n.
	\end{aligned}
	\end{equation}
	Here, we denote by $I(t)$ the random input, modelled as an Ornstein-Uhlenbeck process with mean $\mu$, noise magnitude $\sigma$, and reversion strength $\theta$; by $X_\nu(t)$ the concentration of material in compartment $\nu$; and by $k$ the progression rate of material from one compartment to the next. All parameters are real and positive, and $W(t)$ represents a Wiener process such that $W(t + \eta) - W(t)$ is normally distributed with mean zero and variance $\eta$.
	
	More compactly, we write
	\begin{equation}\label{sys_md-on}
		\dd \Xt = -\bm\Theta \left(\Xt - \bm\mu\right) \, \dd t + \mathbf{S} \, \dd \Wt,
	\end{equation}
	where we define
	\begin{equation}\label{sys_matrices}
		\bm\Theta = \begin{pmatrix}
			\theta & 0  & 0   & \cdots & 0\\
			-1 	   & k  & 0    & \cdots & 0 \\
			0 	   & -k & k   & \cdots & 0 \\
			\multicolumn{3}{c}{\vdots} & \ddots & \\
			0 	   & 0  & 0   & \cdots & k
		\end{pmatrix},\qquad 
		\bm\mu = \mu \begin{pmatrix}1 \\ \bm\Theta_{22}^{-1} \mathbf{e}_1\end{pmatrix},\qquad
 		\mathbf{S} = \begin{pmatrix}
 			\sigma & \mathbf{0} \\
 			\mathbf{0} & \mathbf{0}
 		\end{pmatrix},
	   \end{equation}
	for $\Xt = \big[I(t),X_1(t),\hdots,X_n(t)\big]^\intercal$. For notational convenience, we interchangably refer to $I(t)$ as $X_0(t)$ (i.e., the input is thought of as compartment $\nu = 0$). \Cref{sys_md-on}  demonstrates that the system is a multidimensional Ornstein-Uhlenbeck process and, therefore, a Gaussian process.  We refer to $\bm\Theta$ as the \textit{connectivity matrix}, as it plays a role similar to that in graph and network theory, defining the connectivity between compartments in the system. Therefore, provided the system remains linear, we can arbitrarily express systems with any network structure (i.e., with non-local feedbacks or multiply-connected components) using \cref{sys_md-on}. The form of $\bm\mu$, which contains the lower block matrix $\bm\Theta_{22}$ corresponding to $\bm\Theta$ with the first row and column excluded, simplifies for the system in \cref{fig1} and \cref{sys} to $\big[\mu,\mu/k,\hdots,\mu/k\big]^\intercal$. Unless otherwise stated, we fix $\theta = \mu = k = 1$ and $\sigma = 0.5$ as default parameter values when producing simulation results.
		
	There are various initial conditions that we consider in this work, based on the assumption that a stationary limiting distribution for $\Xt$ exists (since $\theta,k > 0$, this is always true for the form of $\bm\Theta$ expressed above, and more generally provided that all eigenvalues of $\bm\Theta$ have positive real part \cite{Meucci.2009}). The first relevant initial condition is where $\Xz$ is entirely specified. We refer to this choice as the \textit{fixed} initial condition. For the virus-cell lysis problem, we may be interested in setting $\Xz = \big[\mu,0,\hdots,0\big]^\intercal$; i.e., the concentration is zero in all compartments, and the input is initiated at its mean. A second, more biologically realistic initial condition, is where all compartments are initiated with zero concentration, but where the input is initiated from its stationary distribution $I(0) \sim \mathcal{N}(\mu,\sigma^2 / (2\theta))$. We refer to this as the \textit{partially-fixed} initial condition. The final initial condition, of interest given that it greatly simplifies some of the analysis, is where all compartments are initiated from the joint stationary distribution for the system. We refer to this as the \textit{stationary} initial condition and the system as a whole in this case as the \textit{stationary system}.

	\section{Results and Discussion}
	\subsection{Preliminaries}
	
	The multivariate Ornstein-Uhlenbeck process conditioned on the initial condition $\Xz$ has exact solution \cite{Meucci.2009,Vatiwutipong.2019}
		\begin{equation}\label{ou_conddist}
			\Xt | \Xz \sim \mathcal{N}(\mathbf{m}(t),\bm\Sigma(t)),	
		\end{equation}
	where
		\begin{subequations}
		\begin{align}
			\mathbf{m}(t) &= \bm\mu + \mathrm{e}^{-\bm\Theta t}(\Xz - \bm\mu),\label{ou_condexp}\\
			\mathrm{vec}(\bm\Sigma(t)) &= \sigma^2 (\bm\Theta \oplus \bm\Theta)^{-1}\bigg(\mathbf{I} - \mathrm{e}^{-(\bm\Theta \oplus \bm\Theta) t}\bigg) \mathbf{e}_1,\label{ou_condcov}
		\end{align}
		\end{subequations}
	and where $\oplus$ is the Kronecker sum. It follows directly that the stationary distribution, should it exist, is given by
		\begin{equation}\label{stationary_dist}
			\lim_{t\rightarrow\infty}\Xt \sim \mathcal{N}\big(\bm\mu, \bm\Sigma_\infty\big)\quad\text{and}\quad\mathrm{vec}(\bm\Sigma_\infty) = \sigma^2 (\bm\Theta \oplus \bm\Theta)^{-1} \mathbf{e}_1.
		\end{equation}

	We highlight that the non-stationary covariance matrix (\cref{ou_condcov}) does not depend on the initial condition $\Xz$ and that the mean $\mathbf{m}(t)$ is an affine transformation of the initial condition $\Xz$. Therefore, for $\Xz \sim \mathcal{N}(\mathbf{m}_0,\mathbf{\Sigma}_0)$, we have that
		\begin{equation}\label{ou_dist}
			\Xt \sim \mathcal{N}\Big(\bm\mu + \mathrm{e}^{-\bm\Theta t}(\mathbf{m}_0 - \bm\mu), \mathbf{\Sigma}(t) + \mathrm{e}^{-\bm\Theta t} \mathbf{\Sigma}_0 \mathrm{e}^{-\bm\Theta^\intercal t} \Big).
		\end{equation}
	This expression reduces to the fixed initial condition for $\mathbf{\Sigma}_0 = \mathbf{0}$, to the semi-fixed initial condition for $\mathbf{\Sigma}_0 = \mathrm{diag}(\sigma^2 / (2\theta),0,\dots,0)$, and to the stationary initial condition for $\mathbf{\Sigma}_0 = \mathbf{\Sigma}_\infty$.

	The final result for the multivariate Ornstein-Uhlenbeck process that is relevant is the joint distribution of $\mathbf{X}_{t_1,t_2,\dots} = \big[\mathbf{X}(t_1), \mathbf{X}(t_2),\dots\big]^\intercal$, which is multivariate normal with covariance matrix
		\begin{equation}\label{ou_joint_cov}
			\bm\Sigma_{t_1,t_2,\dots} = \begin{pmatrix}
 				\bm\Sigma(t_1)  & \bm\Sigma(t_1) \mathrm{e}^{-\bm\Theta^\intercal (t_2 - t_1)} & \bm\Sigma(t_1) \mathrm{e}^{-\bm\Theta^\intercal (t_3 - t_1)} & \cdots \\
 				\mathrm{e}^{-\bm\Theta (t_2 - t_1)} \bm\Sigma(t_1)  & \bm\Sigma(t_2) & \bm\Sigma(t_2) \mathrm{e}^{-\bm\Theta^\intercal (t_3 - t_2)} & \cdots \\
 				& \vdots & &\ddots \\
 			\end{pmatrix}.
		\end{equation}
	If $t_i \gg 0$ for all $i$ such that $\mathbf{\Sigma}(t_i) = \mathbf{\Sigma}_\infty$, then \cref{ou_joint_cov} corresponds to the joint stationary distribution of $\mathbf{X}_{t_1,t_2,\dots}$ and, along with $\mathbb{E}(\mathbf{X}(t_i)) = \bm\mu$, fully defines the system as a stationary Gaussian process. Furthermore, we can derive the distribution for all initial conditions (i.e., \cref{ou_condcov,stationary_dist,ou_dist}) from the joint stationary distribution from marginalising or conditioning the joint stationary distribution accordingly (this is straightforward for the multivariate normal distribution, see \cite{Eaton.2007}). 
		
	\subsection{Quantifying smoothing in linear multicompartment processes}
		
	The structure of $\mathbf{S}$, whereby noise enters the system only through the first compartment independently of other compartments, results in a simpler form of the stationary covariance matrix, given by \cref{stationary_dist} and which we now denote simply by $\bm\Sigma_\infty$, compared to the standard multivariate Ornstein-Uhlenbeck process. In particular, $\bm\Sigma_\infty$ depends only upon the first column of $(\bm\Theta \oplus \bm\Theta)^{-1}$. In the supplementary material, we provide a full derivation for analytical expressions for $\bm\Sigma_\infty$ in two cases: the first where $\theta = k$, and the second where both $\theta$ and $k$ are allowed to vary freely. In this section, we summarise and discuss the main results.
	
	For $\theta = k$, elements of the symmetric matrix $\bm\Sigma_\infty$ are given by the recurrence relation
		\begin{equation}
			\bm\Sigma_{\infty}^{(i,j)} = \dfrac{\bm\Sigma_{\infty}^{(i,j-1)} + \bm\Sigma_{\infty}^{(i-1,j)}}{2}, \quad i,j = 2,3,\dots
		\end{equation}
	subject to the boundary conditions
		\begin{equation}
			\bm\Sigma_\infty^{(1,1)} = \dfrac{\sigma^2}{2\theta}\quad\text{ and }\quad\bm\Sigma_\infty^{(1,i)} = \dfrac{\bm\Sigma_\infty^{(1,i-1)}}{1 + \theta}, \quad i = 1,2,\dots
		\end{equation}

\begin{figure}[!t]
		\includegraphics[width=\textwidth]{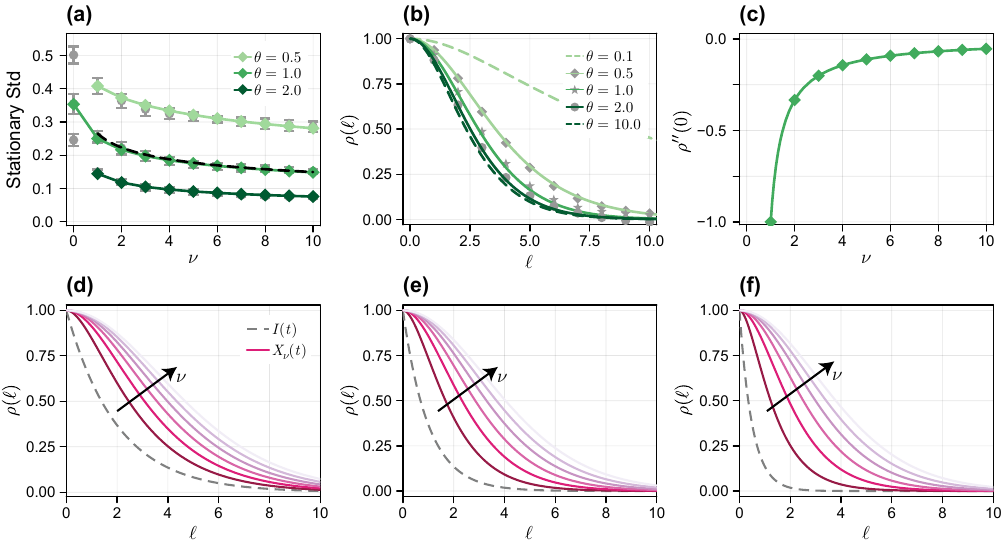}
		\caption[Figure 2]{{\bfseries Variance and autocorrelation function for linear multicompartment model.} (a) Stationary standard deviation as a function of compartment number (compartment $\nu = 0$ refers to $I(t)$). Shown are mean $\pm$ std from numerical simulations constructed from 10 replicates of 100 simulations (grey), and the corresponding analytical solution (colour). (b) Simulated and analytical ACFs for compartment $\nu = 3$. (c) Second derivative of the ACF at $\ell = 0$ for $\theta = k = 1$ calculated directly from \cref{rho_theta1} (diamonds), and using the analytical expression in \cref{acf_curvature} (solid). (d--f) Analytical ACFs for $I(t)$ (grey dashed) and $X_\nu(t)$ (colour of increasing brightness for $\nu = 1,2,\dots,6$). Unless otherwise stated, the other parameters are fixed at $\theta = \mu = k = 1$, and $\sigma = 0.5$.}
		\label{fig2}	
	\end{figure}

	The recurrence relation yields
		\begin{equation}\label{covinf_unity}
			\bm\Sigma_\infty^{(i,j)} = \dfrac{\sigma^2\Gamma(i+j-1)}{2^{i+j-1}\Gamma(i)\Gamma(j)\theta^3},
		\end{equation}
	for covariances relating to the compartments themselves (i.e., $i,j = 2,3,\dots$). Thus, the stationary variance of compartment $\nu \ge 1$ is given by
		\begin{equation}\label{varinf_unity}
			\sigma^2_\nu = \bm\Sigma_\infty^{(\nu+1,\nu+1)} \sim \dfrac{\sigma^2}{2 \theta^3 \sqrt{\nu \pi}},
		\end{equation}
	where we have applied Stirling's approximation \cite{Robbins.1955} to derive an asymptotic expression for the large compartment number, $\nu \gg 1$, limit. In \cref{fig2}a, we compare both the exact and asymptotic expressions for the stationary variance to a numerical approximation produced through repeated simulation of the SDE. Even for $\nu = 1$, the asymptotic expression produces excellent agreement with simulation results.
	
	Relaxing the restriction that $\theta = k$ yields a closed form solution for $\bm\Sigma_\infty$, which simplifies along the diagonal to yield
		\begin{equation}\label{sigma2_nu}
			\sigma_\nu^2 = \dfrac{\sigma^2}{k^2\theta(1 - (\theta / k)^2)^\nu}\dfrac{\mathrm{B}\left(\tfrac{1 - \theta / k}{2},\nu,\nu\right)}{\mathrm{B}(\nu,\nu)},
		\end{equation}
	where $\mathrm{B}(\cdot,\cdot)$ and $\mathrm{B}(\cdot,\cdot,\cdot)$ refer to the Beta function and incomplete Beta function, respectively. We show both analytical and simulation results for $\sigma_\nu^2$ in this more general case in \cref{fig2}a. 

	Taken all together, the results in \cref{fig2}a show that the variance dissipates as the compartment number increases. However, this happens relatively slowly: the analytical expression for $\theta = k$ provides a rate of decay of order $\nu^{-1/2}$. Importantly, the expression in \cref{varinf_unity} indicates that the variance does, indeed, tend to zero as $\nu \rightarrow \infty$. To the best of our knowledge it is not possible to derive a similar expression for general $\theta$, however we provide in the supplementary material a simple proof that $\sigma^2_{\nu+1} < \sigma^2_{\nu}$ for $\theta > 0$ to show that the variance is a strictly decreasing function and in fact tends to zero for large compartment numbers. From this, we also gain insight into the asymptotic behaviour of the solution for small or large $\theta$. As $\theta \rightarrow \infty$, $\sigma_0^2 \rightarrow 0$, and so the system becomes fully deterministic. In the problem itself, this represents a large mean reversion strength, so that effectively $I(t) \equiv \mu \: \forall \: t$. For $\theta = 0$, the input function becomes purely Brownian motion and the stationary solution does not exist. 
			
	While compartment variances and covariances are statistics relating to the process at a single time point, the autocorrelation function (ACF) provides insight into how smooth the resultant time-series is. By deriving an analytical expression for $\mathrm{e}^{-\bm\Theta}$, we can  obtain an analytical expression for the autocorrelation function (ACF). While this is relatively straightforward for the $\theta = k$ case, it is, however, significantly more involved in the more general case, for which the expression obtained is no more helpful that the analytical expression for the autocorrelation function involving the calculation of $\mathrm{e}^{-\bm\Theta}$ directly (supplementary material). For $\theta = k$ we obtain
		\begin{equation}\label{rho_theta1}
			\rho_\nu(\ell) := \mathrm{corr}(X^{(\nu)}_t,X^{(\nu)}_{t+\ell}) = \mathrm{e}^{-k\ell} {}_1F_1(-\nu, -2\nu,2k\ell),
		\end{equation}
	where ${}_1F_1(\,\cdots)$ is the confluent hypergeometric function. \Cref{rho_theta1} can be expressed as
		\begin{equation}\label{rho_theta1-alt}
			\rho_\nu(\ell) = \mathrm{e}^{-k\ell}\left(1 + k\ell + \sum_{i=2}^\nu c_{i,\nu} (k\ell)^i\right),
		\end{equation}
	where coefficients $c_{i,\nu}$ depend on $i$ and $\nu$, which yields a simple expression for $\nu = 1$. We also obtain the scaling of the autocorrelation as a function of $\nu$ by calculating the curvature of the ACF at $\ell = 0$,
		\begin{equation}\label{acf_curvature}
			\rho''_\nu(0) = \dfrac{k^2}{1 - 2\nu},	
		\end{equation}
	where $'$ indicates a derivative with respect to the lag, $\ell$. In \cref{fig2}b, we compare analytical expressions for the ACF to those obtained through simulation, and in \cref{fig2}d--f we show the ACF for the system in \cref{fig1} for $\theta = 0.5$, $1$ and $2$. In \cref{fig2}c we compare the analytical expression for the ACF curvature (\cref{acf_curvature}) to that calculated directly from \cref{rho_theta1} using numerical methods.
	
	As expected, we see qualitatively from the results in \cref{fig2} that further compartments remain correlated for longer. Considered alone, the stationary process $X_\nu(t)$ is itself an, albeit non-Markovian, Gaussian process, fully defined by its variance and autocorrelation function. Therefore, should we normalise each compartment by its respective standard deviation, $X_\nu(t) / \sigma_\nu(t)$, the properties of the resultant process are encoded entirely in the ACF. Given the system in \cref{sys}, we expect $X_\nu(t)$ to be $\nu$-times differentiable (the input, $I(t) = X_0(t)$ is nowhere differentiable) and therefore expect that further compartments will be smoother. For $\nu \ge 1$, we can see such smoothing directly from the ACF curvature in \cref{acf_curvature}. For a small increment, $\ell \ll 1$, we have that $\rho_\nu(\ell) \sim 1 + (\rho_\nu''(0) / 2) \ell^2$ and therefore $\rho_{\nu_1}(\ell) \sim \rho_{\nu_2}(\omega_{\nu_1,\nu_2} \ell)$ where
	\begin{equation}\label{dilation}
		\omega_{\nu_1,\nu_2} = \sqrt{\dfrac{\rho_{\nu_2}''(0)}{\rho_{\nu_1}''(0)}} = \sqrt{\dfrac{1 - 2\nu_1}{1 - 2\nu_2}},
	\end{equation}
	gives the dilation factor. Should we take $\nu_1 = 1$, then $X_{\nu}(t)$ varies a factor of $\omega_{1,\nu} = \sqrt{2\nu - 1}$ more slowly than $X_1(t)$.

	\subsection{Intrinsic noise}

	By modelling the internal dynamics using a system of deterministic ODEs, we have implicitly assumed that the dimensional size of each compartment is sufficiently large relative to the input that we can neglect intrinsic noise. In this section, we relax this assumption and apply the linear noise approximation \cite{vanKampen.2007} (also known as the system-size expansion) to study the relative contribution to the stationary variance from both fluctuations in the input and from intrinsic noise arising for intermediate system sizes. We assume that $\tilde{X}_\nu(t) = V X_\nu(t)$ corresponds to a dimensional concentration of material, where $\tilde{X}_\nu(t) \sim \mathcal{O}(V)$ such that $V$ is the \textit{system size}. 
	
	The governing equations for the internal compartments become
		\begin{equation}
			\dd \tilde{X}_\nu = \Big(k \tilde{X}_{\nu-1}(t) - k \tilde{X}_\nu(t)\Big)\,\dd t + \sqrt{k/V}\Big(\dd W_{\nu-1} - \dd W_\nu\Big),\qquad \nu = 2,\dots,n.
		\end{equation}
	The system can still be viewed as multivariate Ornstein-Uhlenbeck process, although $\mathbf{S}$ (\cref{sys_md-on,sys_matrices}) is no longer a single-element matrix but has elements on both the main and lower diagonal.
	
\begin{figure}[!t]
		\centering
		\includegraphics[scale=1.0]{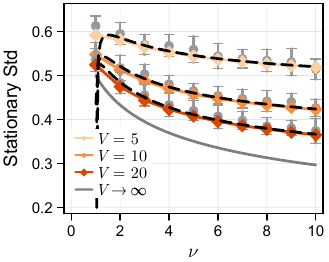}
		\caption[Figure 3]{{\bfseries Adjustment to stationary variance due to intrinsic noise.} Stationary standard deviation as a function of compartment number for (coloured curves) systems of various system sizes. The $V \rightarrow\infty$ limit (grey solid) corresponds to \cref{sigma2_nu}. Also shown are mean $\pm$ std from numerical simulations constructed from 20 replicates of 200 simulations (grey), and asymptotic approximations constructed using Stirling's formula (black dashed). Other parameters are fixed at $\theta = \mu = k = 1$, and $\sigma = 0.5$.}
		\label{fig3}
	\end{figure}

	As fluctuations driven by the Wiener processes $W_\nu$ for $\nu \ge 1$ are independent of fluctuations in the input, we can decompose the stationary variance into contributions each from the input signal and intrinsic noise. The full derivation of the stationary variance for systems with finite $V$ are given as supplementary material. In summary, we stationary variance is now given by
		\begin{equation}\label{sigma_intr}
			\sigma_{\text{intr},\nu}^2 = \underbrace{\vphantom{\Bigg(}\sigma_\nu^2}_{\text{Eq. (\ref{sigma2_nu})}} + \underbrace{\vphantom{\Bigg(}\dfrac{1}{V}\left(1 - \dfrac{\Gamma(2\nu-1)}{2^{2\nu-1} \Gamma(\nu)^2} \right)}_{\text{Intrinsic noise}}.
		\end{equation}
	Of particular note is that the contribution from intrinsic noise is both independent of $k$ and bounded below by a minimum contribution of $1/(2V)$ for $\nu = 1$. In \cref{fig3}, we compare simulation results to both \cref{sigma_intr} and the large system-size case, $V\rightarrow\infty$, which we denote as before by $\sigma_\nu^2$. 
	
	For sufficiently large $\nu$, we can again apply Stirling's approximation \cite{Robbins.1955} to see that the contribution from intrinsic noise behaves like  
		\begin{equation}
			\dfrac{1}{V}\left(1 - \dfrac{1}{2\sqrt{\pi\nu}}\right).
		\end{equation}
	Thus, while $\sigma_\nu^2$ vanishes, $\sigma_{\text{intr},\nu}^2$ tends towards $1 / V$ as $\nu\rightarrow\infty$. Moreover both these limits are approached at rates of order $\nu^{-1/2}$. For all compartments, the contribution to the stationary variance from intrinsic noise is $\mathcal{O}(V^{-1})$, which compares (for the $\theta = k$ case) to the contribution from the input of $\mathcal{O}(\sigma^2\theta^{-3}\nu^{-1/2})$. Therefore, the model that neglects intrinsic noise is valid not only for systems with sufficiently large system sizes, but also for systems with low compartment numbers where $\sigma^2\theta^{-3} \gg V^{-1}$.

	\subsection{First passage time (FPT)}

\begin{figure}[!t]
		\centering
		\includegraphics[scale=1.0]{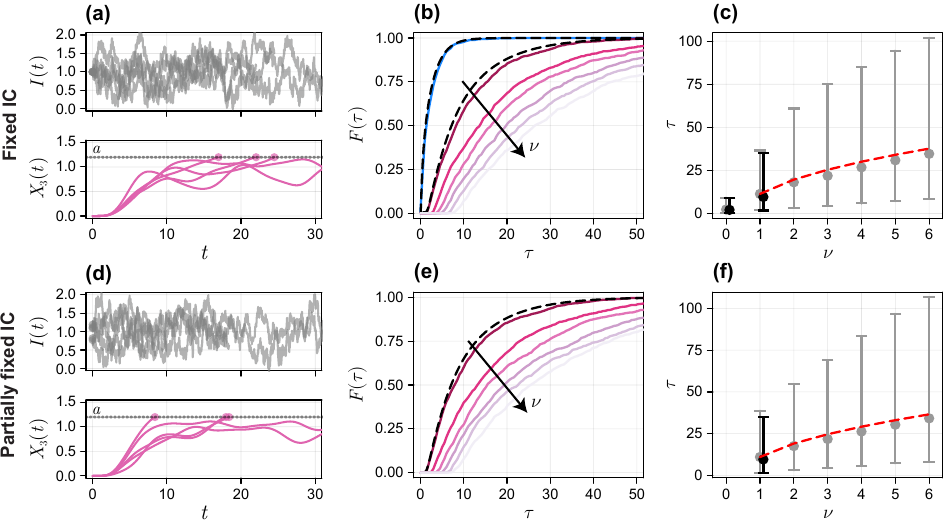}
		\caption[Figure 4]{{\bfseries First passage time distributions for linear multicompartment model.} (a,d) Realisations of a three compartment system initiated using (a) the fixed initial condition and (d) the partially-fixed initial condition. Solutions are terminated at $t = \tau : X_3(\tau) > a$, yielding $\tau$ as the FPT. (b,e) Distribution function for the FPT constructed from (colour) 1000 realisations of the SDE and (dashed black) a finite difference solution to \cref{volterra}. (c,f) Mean, 2.5\% quantile, and 97.5\% quantile for the FPT distribution constructed from (grey) 1000 realisations of the SDE and (black) a finite difference solution to \cref{volterra}. Shown in red-dashed is an approximation to the mean FPT constructed by scaling the FPT for $\nu = 1$ based on matching the second-derivative of the autocorrelation function (\cref{acf_curvature}). The barrier for each compartment is located at $a = \tilde{a}\sigma_\nu$ where $\tilde{a} = 1$. Other parameters are fixed at $\theta = \mu = k = 1$, and $\sigma = 0.5$.}
		\label{fig4}
	\end{figure}

	Motivated by the viral replication problem, we are now interested in studying the FPT distribution of individual compartments. That is, the time at which $X_\nu(t)$ first crosses the threshold value $X_\nu(t) = a$ from below, for $a > X_\nu(0)$. To effectively compare FPT distributions between compartments, we scale the threshold by the associated stationary standard deviation of the relevant compartment such that $a = \mu + \tilde{a} \sigma_\nu$, where $\sigma_\nu$ is given by \cref{varinf_unity} and $\tilde{a}$ is specified. Formally, we define the FPT by $\tau = \inf\{t : X_\nu(t) \ge a\}$ and its associated probability density and distribution functions by $f(\tau)$ and $F(\tau)$, respectively. We focus on fixed and partially-fixed initial conditions where we ensure that $X_\nu(0) < a$, demonstrated in \cref{fig4}a,d, respectively, for $\nu = 3$ and $\tilde{a} = 1$. 
	
	It is not generally possible to derive an analytical expression for $f(t)$, nor to formulate a closed-form integral equation that can be solved numerically. The only general way to solve for $f(t)$ is through a numerical solution to the Fokker-Planck equation, a $\nu+1$ dimensional partial differential equation. However, following the procedure for Markovian Gaussian processes \cite{Bernard.1972,Grigoriu.2020} we are able to formulate a reasonable approximation for $\nu = 1$, although we revert to simu lating the FPT for the more general case.
	
	Denote by $p_\nu(x,t)$ the density of the random variable $X_\nu(t)$, and by $p_\nu(x,t,\tau)$ the joint density with the first passage time. By marginalising $p_\nu(x,t,\tau)$ with respect to $\tau$ we have that
		\begin{equation}\label{volterra}
			p_\nu(a,t) = \int_0^\infty p_\nu(a,t,\tau) \, \dd \tau = \int_0^t K(t,\tau) f(\tau) \, \dd\tau,
		\end{equation}
	where $K(t,\tau) = p_\nu(a,t|\tau)$ is the density of $X(t)$ given the first passage time $\tau$. The upper limit of $t$ in the second integral arises since $p_\nu(a,t|\tau) = 0$ for all $t < \tau$; in other words, should a passage not have occurred by time $t$, then $X_\nu(t) < a$ and so the particle cannot be at location $X_\nu(t) = a$. \Cref{volterra} is a Volterra equation of the first kind and, while generally difficult, can be solved numerically provided $K(t,\tau)$ can be computed. As $I(t) = X_0(t)$ is Markovian, $p_0(a,t|\tau) = p_0(a,t|a,\tau)$ is readily available and for certain parameter combinations, \cref{volterra} can be solved analytically to give the FPT density of the Ornstein-Uhlenbeck process \cite{Mehr.1965,Ricciardi.1988}.
		
	For $\nu > 0$ the conditional probability $p_\nu(a,t|\tau)$ cannot be calculated exactly; further we find that the standard approach of approximating $p_\nu(a,t|\tau) \approx p_\nu(a,t|a,\tau)$ \cite{Bernard.1972} provides relatively poor results. To obtain a more accurate approximation, we note that the full state process $\Xt$ is Markovian, and that $X_\nu(s) = a$ and $X_\nu'(s) > 0$ if and only if $s$ is a passage time. By \cref{sys}, $X_\nu'(s)$ is a linear combination of other states, and so the random variable $\big[X_\nu(s),X_\nu'(s)\big]$ is Markovian with a multivariate normal distribution. Thus, we approximate
		\begin{equation}\label{approx_kernel}
			p_\nu(a,t|\tau) \approx p_\nu(a,t|X(\tau) = a, X'(\tau) > 0).
		\end{equation}
	Note that \cref{approx_kernel} is not exact despite the full state being Markovian as we have not conditioned on a point, but rather the range $X'(\tau) > 0$; while the distribution of $X'(s)$ is normal, the distribution of $X'(\tau)$ is not necessarily so. In future, additional approximations based on so-called FPT functionals could potentially be constructed \cite{Singh.2022sq}. We find that a numerical solution to \cref{volterra,approx_kernel} gives a reasonable approximation to $f(\tau)$ for $\nu = 1$ (\cref{fig4}b). 
		
	Results in \cref{fig4}b,e show the FPT distribution function, $F(\tau)$, for both the fully- and partially-fixed initial conditions for $\theta = k = 1$, respectively. The coloured curves are produced from 1000 realisations of the SDE model, and the black curves from a numerical solution to \cref{volterra}. An interpretation of $S(\tau) = 1 - F(\tau)$ is that of the \textit{survival probability}: should a passage indicate system failure, $S(\tau)$ represents the probability that a system is functional at time $\tau$. For the virus-cell lysis problem, we interpret this as the probability that cell lysis has not occured, and viral production is ongoing. Visual inspection of the results in \cref{fig4} reveal little difference between both initial conditions, particularly for larger compartment numbers. This observation is unsurprising upon comparison between the magnitude of the mean FPT, $\mathbb{E}(\tau) \sim \mathcal{O}(10)$, and the largest eigenvalue of $-\bm\Theta$, $\lambda = -1$, demonstrating that the influence of the initial condition decays like $\mathrm{exp}(-t)$ (\cref{ou_condexp}), much faster than the mean FPT.

	The most obvious result from \cref{fig4}b,e, as one might expect from the analysis of compartment smoothing in the previous section, is that the FPT is generally larger for further compartments; accounting for differences in the stationary variance by comparing compartments across the same value of $\tilde{a}$ indicates that further compartments can be thought to be more robust to external noise. Not only does the expected FPT increase (equal to the area under the survival function $S(\tau)$), but so too do the lower quantiles, evidenced by the time taken for the distribution function $F(\tau)$ to visually become non-zero. We investigate these qualitative observations further in \cref{fig4}c,f, by calculating the mean, 2.5\% and 97.5\% quantiles for the FPT for each compartment. Aside from an overall increase in the FPT, there is a notable increase in the inter-quantile range or variance for larger compartment numbers.
	
	By normalising the location of the threshold by the compartment standard deviation, the FPT is almost entirely a function of the ACF for each compartment. The compartment variance still plays a role for the initial conditions considered in \cref{fig4}, as the relative distance between the initial condition $X_\nu(0) = 0$ and the threshold depends on $\sigma_\nu$. The settling phase, however, occurs relatively quickly: for $k = 1$, the system settles to equilibrium like $\exp(-t)$, much faster than the typical FPT. Thus, the compartment smoothing effect characterised by $\rho_\nu(\ell)$, which provides the temporal scale, is the primary factor that drives increases in FPT for further compartments. In \cref{fig4}c,f, we show that the small-lag ACF dilation factor, given analytically by \cref{dilation}, provides an excellent match to numerical results for the mean first passage time, confirming qualitatively that, similarly to the compartment variance, the FPT scales like $\sqrt{\nu}$. Additional results (supplementary material), constructed for various values of $k$ and hence various ACFs, demonstrate strong correlation (Spearman correlation of $-0.987$) between the ACF curvature and the expected FPT.

	\subsection{Production before system failure}

	While for many systems the FPT, $\tau$, may itself be of primary interest, for others it may be a so-called FPT functional that is important \cite{Singh.2022sq}. For the virus-cell lysis problem, for example, of primary interest may be the total amount of virus produced by the system prior to cell lysis: a viral genotype that maximises per-host-cell virion production may have a fitness advantage over others. Also seen in the virus literature is that this so-called ``burst-size'' declines with cell size \cite{DeLong.2022}, potentially linking the structure of a particular viral replication network to an optimal cell size and, by extension, an optimal cell type.
	
\begin{figure}[!t]
		\centering
		\includegraphics[width=\textwidth]{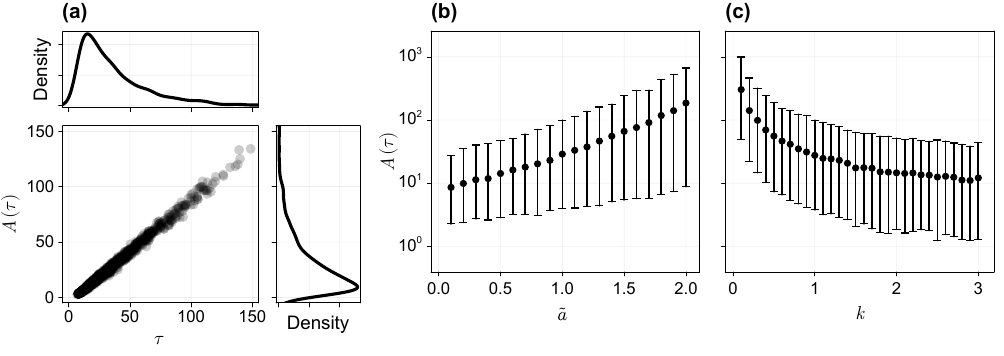}
		\caption[Figure 5]{{\bf Cumulative production at the first passage time.} We investigate the cumulative production at the first passage time, denoted by $A(\tau) = \int_0^\tau k X_6(t)\, \dd t$, for a six compartment system. (a) Relationship between $\tau$ and $A(\tau)$ based on 1000 replicates of the SDE with $\tilde{a} = k = 1$. We show both a scatter plot of the joint density and kernel density estimates constructed from the marginals. (b,c) Mean, 2.5\% quantile, and 97.5\% quantile of $A(\tau)$ constructed from 1000 replicates of the SDE as (b) $\tilde{a}$ is varied, and (c) $k$ is varied.}
		\label{fig5}	
	\end{figure}

	We define cumulative production as the total amount of material to pass out of the system, $A(t) = \int_0^t k X_\nu(t) \,\dd t$, and investigate $A(\tau)$ through simulation in \cref{fig5} for $\nu = 6$ and $\theta = k = 1$. Results in \cref{fig5}a show that $A(\tau)$ is highly correlated with $\tau$. We expect this, as both initial settling and ACF decay occur much faster on average than $\tau$. Thus, for the linear system considered in this work, we hypothesise that a genotype that maximises the expected FPT could be considered equivalent to one that maximises the production prior to lysis.
	
	In \cref{fig5}b,c we perform a parameter sweep to determine the distribution of $A(\tau)$ as a function of the threshold location, $\tilde{a}$, and the compartment transfer rate, $k$. Clearly, for thresholds that are larger, and crossed more infrequently, we see an increase in $A(\tau)$. We view the threshold location as a feature of the host-system: for the virus-cell lysis problem, this is a biological feature of the host cell, and not a feature of the viral genome. Of direct interest is the relationship between $k$ and $A(\tau)$. The results in \cref{fig5}c show that systems with large compartment transfer rates have lower cumulative production than those that operate slowly. While we cannot interpret the expression for the ACF curvature (\cref{acf_curvature}) exactly for $\theta \neq k$, the expression does suggest that the ACF curvature is proportional to $k^2$, thus increasing the rate at which material travels through the system is detrimental to robustness.
	
	In the virus-cell lysis problem, these results appear to suggest that production can be maximised should the system tune itself to operate slowly (i.e., small values of $k$). However, this observation ignores potential tradeoffs induced by immune-response mechanisms in the host \cite{Wodarz.2019}: while production in an isolated or experimental system might be maximised by slowly material progression, operating quickly carries advantages of lysis occuring due to material production, and not through an immune response of the host organism. Analysis of more complicated, non-linear, compartment processes may also yield non-monotonic relationships between fitness and speed or other parameters; such analysis is, however, beyond the scope of the present work.

	\subsection{Non-local feedback to alter system robustness}

	Another way in which systems can potentially increase their robustness to input noise is through non-local feedback or feedforward loops. Such a phenomenon, where progression through the virus life cycle is not unidirectional, is common with evidence in the virus literature: for example, the complex network-like replication cycle seem in the human ademovirus \cite{Flint.2015}. Hepatitus B, an enveloped DNS virus, recycles new viral DNA that is awaiting repackaging back into the nucleus \cite{Jayalakshmi.2013,Saraceni.2021}: analogous, in our simple linear model, to a feedback from the final to an early compartment. A similar late to early-stage feedback is seen in positive-strand RNA viruses in which newly replicated RNA strands are either encapsidated or reutilised in translation and replication \cite{Nagy.2012}. More generally, positive- and negative-strand RNA viruses have very similar replication structures and an identical feedback mechanism, however are in part distinguished by their feedforward structures \cite{Li.2013,Lundstrom.2021}. Less concretely, replication of the measles virus comprises several feedback and feedforward mechanisms, particularly during the transition and polymerase stages \cite{Su.2021}. 
	
\begin{figure}
		\centering
		\includegraphics{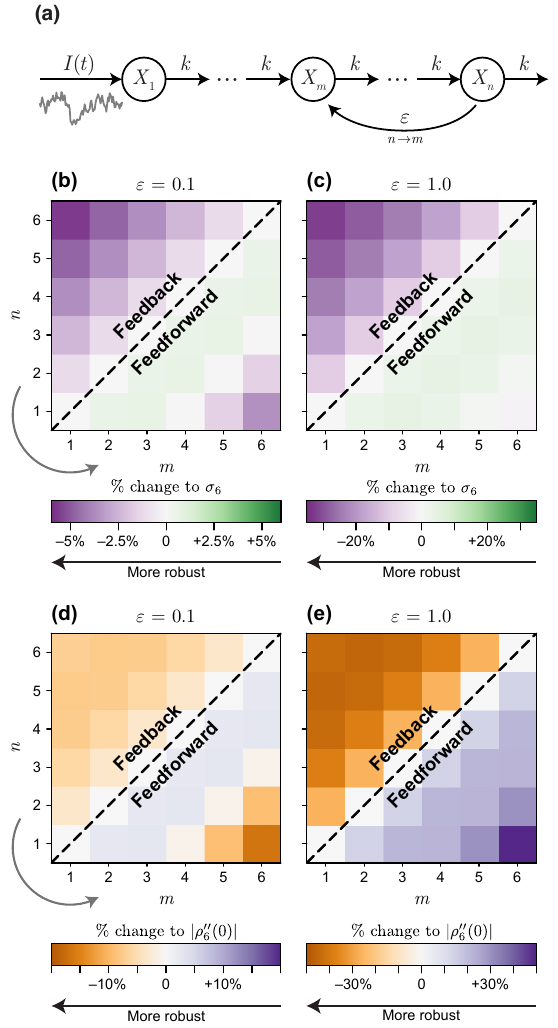}
		\caption[Figure 6]{{\bf System with non-local feedback.} We investigate the effect of a non-local feedback (or feedforward) of magnitude $\varepsilon$ from compartment $n$ to compartment $m$ on the stationary standard deviation of $X_6(t)$, denoted $\sigma_6$, and the magnitude of the ACF curvature, denoted $|\rho''_6(0)|$. In all cases, a reduction in each statistic corresponds to a potentially more robust system.}
		\label{fig6}	
	\end{figure}

	In this section, we investigate the relative change to the stationary variance and ACF curvature in the final compartment of a system with an additional transfer from compartment $n$ to compartment $m$ of magnitude $\varepsilon$ (\cref{fig6}a). A transfer with $m < n$ is considered a feedback, with $m > n$ a feedforward. Since the within system dynamics are deterministic, a transfer with $m = n$ has no net effect on the dynamics.
	
	The results in \cref{fig6}b,c show that the output variance can be reduced, potentially significantly, through a feedback. For the system considered, a feedback from the final to the first compartment has the largest effect, yielding a reduction of over 30\% to the stationary standard deviation. In \cref{fig6}d,e we show a similar affect on the curvature of the ACF. Interestingly, feedbacks from the second last compartment to the first compartment, or the last to the second, yield the greatest reduction in the curvature of the ACF. We also observe a non-linear relationship between the magnitude of the feedback and the ACF curvature. For example, a small ($\varepsilon = 0.1$) feedforward rate from the first to the last compartment yields a reduction in the ACF curvature whereas a large ($\varepsilon = 1$) feedforward rate yields an increase in the curvature.

	\subsection{Continuum limit}

\begin{figure}[!b]
		\centering
		\includegraphics[width=\textwidth]{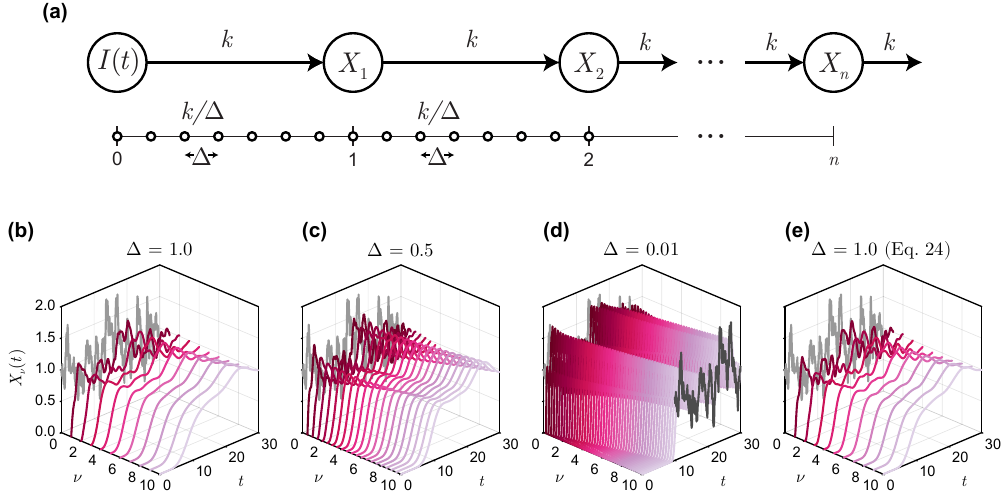}
		\caption[Figure 7]{{\bf Material transport through a system approaching the continuum limit}.  (a) Compartments in the discrete system are divided into subcompartments of ``length'' $\Delta$, while the ``transfer density'' is kept fixed. (b--d) Solutions of the discrete system for decreasing $\Delta$. All systems are subject to identical input $I(t)$ (grey). In (d), the shifted input $I(t - 10)$ is shown at $\nu = 10$ for comparison with the transported concentrations. Other parameters are fixed $\theta = \mu = k = 1$, and $\sigma = 0.5$. (e) Solution to the continuum limit approximation (\cref{continuum3}) for $\Delta = 1$. }
		\label{fig7}	
	\end{figure}

	While the focus of the present work is on multicompartment processes, a natural extension is to investigate smoothing in processes that occur on a continuum (for example, where compartment number, $\nu$, is a continuous measure of how far a particle has progressed through a system). Therefore, we consider a refinement of the discrete process by dividing each compartment into $1 / \Delta$ subcompartments, each of width $\Delta$ (\cref{fig7}a). To maintain the effective total time a particle spends in the system, we assume that the subcompartment transfer rate becomes $\hat{k} = k / \Delta$ such that
	\begin{equation}\label{sys_mod}
	\begin{aligned}
		X_0(t) &= I(t),\\
		\dv{X_i}{t} &= \dfrac{k}{\Delta}\big(X_{i-1}(t) - X_i(t)\big), \quad i > 0,
	\end{aligned}
	\end{equation}
	where $\nu = (i-1)\Delta$. Note that to formulate the input as a boundary condition, we have modified the original system such that the input transfers to the first compartment at rate $k$ (i.e., $X_1(t)$ now experiences an input of $kI(t)$ compared to the input of $I(t)$ in \cref{sys}). This formulation is equivalent to the original formulation for $k = 1$.
	
	We denote $x(\nu,t) = X_i(t)$, and take $\Delta \rightarrow 0$ to yield an advection equation with Dirichlet boundary condition
	\begin{equation}\label{continuum_limit}
	\begin{aligned}
		\pdv{x}{t} &= -k\pdv{x}{\nu},\\
			x(0,t) &= \left\{\begin{array}{ll}
						I(t), & t > 0, \\
						0, & t \le 0, \end{array}\right.\\
			x(\nu,0) &= 0,\qquad \nu > 0.
	\end{aligned}
	\end{equation}
	with exact solution
	\begin{equation}\label{continuum_solution}
		x(\nu,t) = I(t - \nu / k).
	\end{equation}
	As an advection equation, this continuum analogue of the multicompartment system corresponds to exact (undamped) transport of material through the system. We conclude, therefore, that the smoothing we see is a uniquely discrete effect. This conclusion becomes obvious should each compartment be viewed as a well-mixed segment of ``length'' $\Delta$ in $\nu$-space. Transfer between each compartment represents flow across the left boundary. As the ``length'' of each compartment becomes smaller, the left boundary approaches the right, and thus the finite difference between successive compartments diminishes. Thus, smoothing can be viewed as a consequence of within-compartment mixing, a feature of the discrete system that vanishes as the size of each compartment tends to zero.
		
	We perform a numerical experiment in \cref{fig7}b--d and simulate three systems subject to an identical input, with identical total length $n = 10$, and with compartment spacing reducing from $\Delta = 1$ to $\Delta = 0.01$. In effect, the solution to the discrete system corresponds exactly to a forward difference approximation to that of the advection equation (\cref{continuum_limit}). Smoothing in both the autocorrelation and variance is evident in all cases with finite $\Delta$, however, it diminishes significantly for $\Delta = 0.01$.
	
	The advection equation given in \cref{continuum_limit} gives little insight into the behaviour as $\Delta \rightarrow 0$. To derive a continuum limit approximation that captures the smoothing effects in systems with small but finite $\Delta$, we apply the method of multiple scales and choose a `slow' scale of $s = \sqrt{\Delta} (i-1) - kt / \sqrt{\Delta} \sim \mathcal{O}(1)$ (the fast scale, considered in earlier analysis, is $\nu = (i-1)\Delta$). This scaling yields
		\begin{equation}\label{continuum2}
			\pdv{x}{t} = \dfrac{k}{2} \pdv[2]{x}{s},	
		\end{equation}
	or, in the original variables,
		\begin{equation}\label{continuum3}
			\pdv{x}{t} = -k\pdv{x}{\nu} + \dfrac{k\Delta}{2} \pdv[2]{x}{\nu}.
		\end{equation}
	Full details relating to the derivation of this high-order continuum limit approximation are given as supplementary material.
	
	In \cref{fig7}e, we provide a numerical solution to \cref{continuum3} for $\Delta = 1$, showing that this second continuum approximation captures the smoothing effect seen in the discrete model. The inclusion of the diffusion term with coefficient $\mathcal{O}(\Delta)$ demonstrates that, while the problem is singular, smoothing is always present in the discrete system. This observation is, in fact, a well known feature of the discrete system when viewed as an approximate numerical solution to the advection equation with an upwind spatial scheme. Such a scheme can be simply seen to introduce \textit{artificial diffusion}---effectively, smoothing---with form identical to that given by the multiple scales approach in \cref{continuum3}.

\section{Conclusion}

Multicompartment processes are ubiquitous in biology; from linear progression through the cell cycle, to phage replication in bacteria and the propagation of viruses by hijacked cellular machinery. Our analysis demonstrates that even a fundamental linear multicompartment structure provides potential advantages and benefits to the systems that employ them. These results parallel filters and autoregressive models in control theory that allow engineers to control and exploit systems subject to noisy input \cite{Box.2015,Vecchio.2016l57,Khammash.2021}. 

Most notable is the effect of such systems to smooth and provide an additional degree of control over external noise, consequentially increasing resilience and robustness. The inclusion of feedback and feedforward loops can enhance this effect, providing systems with additional degrees of control and contributing to so-called perfect adaptation \cite{Reeves.2019,Khammash.2021}. Such loops provide a potential explanation for out-of-order progression in some systems, for example, whereby viral replication does not occur as a unidirectional process \cite{Louten.2016}. Our work demonstrates that feedback loops could yield a fitness advantage through more favourable statistical properties of viral load compared with perfect progression through the replication cycle. Such results potentially explain the complexity in the replication network structure seen in some viruses; for example, in the human adenovirus \cite{Flint.2015}. Indirectly, these loops (and by extension, more complicated network structures) provide systems with the ability to tune the first passage time distribution, potentially yielding an \textit{optimal} lysis time \cite{Ghusinga.2017,Kannoly.2022df}. While we restrict our analysis to a single non-local feedback or feedforward, future work may study more general optimal network structures informed by more specific biological problems.

Analysis of a linear model system, subject to Ornstein-Uhlenbeck-type noise, allows us to present closed-form expressions for key statistics, providing a fundamental understanding that would be otherwise unavailable for more complicated systems. We reveal that noise dissipates in systems with unidirectional flow, eventually vanishing in an infinite-compartment system. However, we show that this effect is tied to a finite flow rate: scaling to yield continuous flow through a continuum limit approximation reveals that smoothing is a discrete effect, caused by \textit{within compartment} mixing. While many simple results are only available for the constant flow rate assumption, arbitrarily connected linear systems yield a multidimensional Ornstein-Uhlenbeck process with statistical properties computable semi-analytically through the exponentiation of a connectivity matrix. 

Importantly, we lay the foundation for future work to explore the properties of more general multi-compartment systems subject to external noise. For instance, study of the interaction between the external timescales (i.e., autocorrelation of the external input) and the internal timescales (i.e., progression through the system) is highly relevant to specific biological problems: in the virus-cell lysis problem, this interaction is relevant for immune system detection and, therefore, infection clearance. Aside from external noise, other stochastic features, including both between-cell and between-virion heterogeneity and fluctuations in the replication process itself are known to play an important role in within-host virus replication \cite{Timm.2012,Jenner.20181ns,Jones.2021,Sazonov.2022}. Despite these observations, the study of multicompartment problems with the stochastic mathematical models requisite to capture important features is presently scarce, albeit a rich area for both mathematical and biological insight.

\section*{Data availability}
	Code used to produce the results are available on GitHub at \url{https://github.com/ap-browning/multicompartment}.

\section*{Author contributions}
	All authors conceived the study, provided feedback on drafts, and gave approval for final publication. A.P.B. implemented the computational algorithms and drafted the manuscript.

\section*{Acknowledgements}
	The authors thank Ben Hambly, Hilary Hunt, and Mohit Dalwadi for helpful discussions.

	{\footnotesize

\begin{thebibliography}{10}

\bibitem{Cann.2008}
Cann AJ.
\newblock {Replication of viruses}.
\newblock In: Encyclopedia of Virology (Third Edition). Academic Press; 2008.
  p. 406--412.

\bibitem{Louten.2016}
Louten J.
\newblock {Virus Replication}.
\newblock In: Essential Human Virology. Academic Press; 2016. p. 49--70.

\bibitem{Sazonov.2022}
Sazonov I, Grebennikov D, Meyerhans A, Bocharov G.
\newblock {Sensitivity of SARS-CoV-2 life cycle to IFN effects and ACE2 binding
  unveiled with a stochastic model}.
\newblock Viruses. 2022;14(2):403.
\newblock doi:{10.3390/v14020403}.

\bibitem{Campbell.2003}
Campbell A.
\newblock {The future of bacteriophage biology}.
\newblock Nature Reviews Genetics. 2003;4(6):471--477.
\newblock doi:{10.1038/nrg1089}.

\bibitem{Morgan.2007}
Morgan DO.
\newblock {The Cell Cycle: Principles of Control}.
\newblock London: New Science Press Ltd; 2007.

\bibitem{Huang:1996}
Huang CY, Ferrell JE.
\newblock {Ultrasensitivity in the mitogen-activated protein kinase cascade.}
\newblock Proceedings of the National Academy of Sciences.
  1996;93(19):10078--10083.
\newblock doi:{10.1073/pnas.93.19.10078}.

\bibitem{Carr.2019df}
Carr EJ, Simpson MJ.
\newblock {New homogenization approaches for stochastic transport through
  heterogeneous media}.
\newblock The Journal of Chemical Physics. 2019;150(4):044104.
\newblock doi:{10.1063/1.5067290}.

\bibitem{Liu.2004}
Liu L, Liu X, Yao DD.
\newblock {Analysis and optimization of a multistage inventory-queue system}.
\newblock Management Science. 2004;50(3):365--380.
\newblock doi:{10.1287/mnsc.1030.0196}.

\bibitem{Schulte.2014}
Schulte MB, Andino R.
\newblock {Single-cell analysis uncovers extensive biological noise in
  poliovirus replication}.
\newblock Journal of Virology. 2014;88(11):6205--6212.
\newblock doi:{10.1128/jvi.03539-13}.

\bibitem{Jones.2021}
Jones JE, Sage VL, Lakdawala SS.
\newblock {Viral and host heterogeneity and their effects on the viral life
  cycle}.
\newblock Nature Reviews Microbiology. 2021;19(4):272--282.
\newblock doi:{10.1038/s41579-020-00449-9}.

\bibitem{Heaton.2017}
Heaton NS.
\newblock {Revisiting the concept of a cytopathic viral infection}.
\newblock PLoS Pathogens. 2017;13(7):e1006409.
\newblock doi:{10.1371/journal.ppat.1006409}.

\bibitem{Redner.2001}
Redner S.
\newblock {A Guide To First-Passage Processes}; 2001.

\bibitem{Kannoly.2022df}
Kannoly S, Singh A, Dennehy JJ.
\newblock {An optimal lysis time maximizes bacteriophage fitness in
  quasi-continuous culture}.
\newblock mBio. 2022;13(3):e03593--21.
\newblock doi:{10.1128/mbio.03593-21}.

\bibitem{Allen:2011}
Allen LJS.
\newblock {An Introduction to Stochastic Processes with Applications to
  Biology}.
\newblock Boca Raton, Florida: Chapman \& Hall/CRC Press; 2011.

\bibitem{Simpson.2015}
Simpson MJ, Sharp JA, Baker RE.
\newblock {Survival probability for a diffusive process on a growing domain}.
\newblock Physical Review E. 2015;91(4):042701.
\newblock doi:{10.1103/physreve.91.042701}.

\bibitem{Rijal.2022}
Rijal K, Prasad A, Singh A, Das D.
\newblock {Exact distribution of threshold crossing times for protein
  concentrations: implication for biological timekeeping}.
\newblock Physical Review Letters. 2022;128(4):048101.
\newblock doi:{10.1103/physrevlett.128.048101}.

\bibitem{Bernard.1972}
Bernard MC, Shipley JW.
\newblock {The first passage problem for stationary random structural
  vibration}.
\newblock Journal of Sound and Vibration. 1972;24(1):121--132.
\newblock doi:{10.1016/0022-460x(72)90128-9}.

\bibitem{Grigoriu.2020}
Grigoriu M.
\newblock {First passage times for Gaussian processes by Slepian models}.
\newblock Probabilistic Engineering Mechanics. 2020;61:103086.
\newblock doi:{10.1016/j.probengmech.2020.103086}.

\bibitem{Matis.1979}
Matis JH, Wehrly TE.
\newblock {Stochastic models of compartmental systems}.
\newblock Biometrics. 1979;35(1):199.
\newblock doi:{10.2307/2529945}.

\bibitem{Yu.2004}
Yu J, Wehrly TE.
\newblock {An approach to the residence time distribution for stochastic
  multi-compartment models}.
\newblock Mathematical Biosciences. 2004;191(2):185--205.
\newblock doi:{10.1016/j.mbs.2004.06.005}.

\bibitem{Donnet.2013}
Donnet S, Samson A.
\newblock {A review on estimation of stochastic differential equations for
  pharmacokinetic/pharmacodynamic models}.
\newblock Advanced Drug Delivery Reviews. 2013;65(7):929--939.
\newblock doi:{10.1016/j.addr.2013.03.005}.

\bibitem{Mistry.2018}
Mistry BA, D’Orsogna MR, Chou T.
\newblock {The effects of statistical multiplicity of infection on virus
  quantification and infectivity assays}.
\newblock Biophysical Journal. 2018;114(12):2974--2985.
\newblock doi:{10.1016/j.bpj.2018.05.005}.

\bibitem{Teufel.2020}
Teufel AI, Liu W, Draghi JA, Cameron CE, Wilke CO.
\newblock {Uncovering modeling features of viral replication dynamics from
  high-throughput single-cell virology experiments}.
\newblock bioRxiv. 2020; p. 2020.07.09.195925.
\newblock doi:{10.1101/2020.07.09.195925}.

\bibitem{Anderson.2007}
Anderson DF, Mattingly JC, Nijhout HF, Reed MC.
\newblock {Propagation of fluctuations in biochemical systems, I: Linear SSC
  networks}.
\newblock Bulletin of Mathematical Biology. 2007;69(6):1791--1813.
\newblock doi:{10.1007/s11538-007-9192-2}.

\bibitem{Box.2015}
Box GEP.
\newblock {Time series analysis : forecasting and control }.
\newblock Fifth edition. ed. Hoboken, New Jersey: Wiley; 2015.

\bibitem{Meucci.2009}
Meucci A.
\newblock {Review of Statistical Arbitrage, Cointegration, and Multivariate
  Ornstein-Uhlenbeck}.
\newblock SSRN Electronic Journal. 2009;doi:{10.2139/ssrn.1404905}.

\bibitem{Vatiwutipong.2019}
Vatiwutipong P, Phewchean N.
\newblock {Alternative way to derive the distribution of the multivariate
  Ornstein-Uhlenbeck process}.
\newblock Advances in Difference Equations. 2019;2019(1):276.
\newblock doi:{10.1186/s13662-019-2214-1}.

\bibitem{Gardiner}
Gardiner CW.
\newblock {Stochastic Methods}.
\newblock 4th ed. Berlin: Springer; 2009.

\bibitem{Eaton.2007}
Eaton ML.
\newblock {Multivariate Statistics: A Vector Space Approach}.
\newblock Lecture notes-monograph series. 2007;53:1--512.

\bibitem{Robbins.1955}
Robbins H.
\newblock {A remark on Stirling's formula}.
\newblock The American Mathematical Monthly. 1955;62(1):26--29.
\newblock doi:{10.2307/2308012}.

\bibitem{vanKampen.2007}
van Kampen NG.
\newblock {Stochastic processes in physics and chemistry}.
\newblock 3rd ed. Amsterdam: Elsevier; 2007.

\bibitem{Mehr.1965}
Mehr CB, McFadden JA.
\newblock {Certain properties of Gaussian processes and their first-passage
  times}.
\newblock Journal of the Royal Statistical Society: Series B (Methodological).
  1965;27(3):505--522.
\newblock doi:{10.1111/j.2517-6161.1965.tb00611.x}.

\bibitem{Ricciardi.1988}
Ricciardi LM, Sato S.
\newblock {First-passage-time density and moments of the Ornstein-Uhlenbeck
  process}.
\newblock Journal of Applied Probability. 1988;25(1):43--57.
\newblock doi:{10.2307/3214232}.

\bibitem{Singh.2022sq}
Singh P, Pal A.
\newblock {First-passage Brownian functionals with stochastic resetting}.
\newblock Journal of Physics A: Mathematical and Theoretical.
  2022;55(23):234001.
\newblock doi:{10.1088/1751-8121/ac677c}.

\bibitem{DeLong.2022}
DeLong JP, Al-Sammak MA, Al-Ameeli ZT, Dunigan DD, Edwards KF, Fuhrmann JJ,
  et~al.
\newblock {Towards an integrative view of virus phenotypes}.
\newblock Nature Reviews Microbiology. 2022;20(2):83--94.
\newblock doi:{10.1038/s41579-021-00612-w}.

\bibitem{Wodarz.2019}
Wodarz D, Levy DN, Komarova NL.
\newblock {Multiple infection of cells changes the dynamics of basic viral
  evolutionary processes}.
\newblock Evolution Letters. 2019;3(1):104--115.
\newblock doi:{10.1002/evl3.95}.

\bibitem{Flint.2015}
Flint SJ, Racaniello VRVR, Rall GF, Skalka AM, Enquist LWLW.
\newblock {Principles of virology }.
\newblock 4th ed. Washington, DC: ASM Press; 2015.

\bibitem{Jayalakshmi.2013}
Jayalakshmi M, Kalyanaraman N, Pitchappan R.
\newblock {Viral Replication}. 2013;doi:{10.5772/53818}.

\bibitem{Saraceni.2021}
Saraceni C, Birk J.
\newblock {A review of hepatitis B virus and hepatitis C virus
  immunopathogenesis}.
\newblock Journal of Clinical and Translational Hepatology. 2021;9(3):409--418.
\newblock doi:{10.14218/jcth.2020.00095}.

\bibitem{Nagy.2012}
Nagy PD, Pogany J.
\newblock {The dependence of viral RNA replication on co-opted host factors}.
\newblock Nature Reviews Microbiology. 2012;10(2):137--149.
\newblock doi:{10.1038/nrmicro2692}.

\bibitem{Li.2013}
Li D, Wei T, Abbott CM, Harrich D.
\newblock {The unexpected roles of eukaryotic translation elongation factors in
  RNA virus replication and pathogenesis}.
\newblock Microbiology and Molecular Biology Reviews. 2013;77(2):253--266.
\newblock doi:{10.1128/mmbr.00059-12}.

\bibitem{Lundstrom.2021}
Lundstrom K.
\newblock {Self-replicating RNA viruses for vaccine development against
  infectious diseases and cancer}.
\newblock Vaccines. 2021;9(10):1187.
\newblock doi:{10.3390/vaccines9101187}.

\bibitem{Su.2021}
Su JM, Wilson MZ, Samuel CE, Ma D.
\newblock {Formation and function of liquid-like viral factories in
  negative-sense single-stranded RNA virus infections}.
\newblock Viruses. 2021;13(1):126.
\newblock doi:{10.3390/v13010126}.

\bibitem{Vecchio.2016l57}
Vecchio DD, Dy AJ, Qian Y.
\newblock {Control theory meets synthetic biology}.
\newblock Journal of The Royal Society Interface. 2016;13(120):20160380.
\newblock doi:{10.1098/rsif.2016.0380}.

\bibitem{Khammash.2021}
Khammash MH.
\newblock {Perfect adaptation in biology}.
\newblock Cell Systems. 2021;12(6):509--521.
\newblock doi:{10.1016/j.cels.2021.05.020}.

\bibitem{Reeves.2019}
Reeves GT.
\newblock {The engineering principles of combining a transcriptional incoherent
  feedforward loop with negative feedback}.
\newblock Journal of Biological Engineering. 2019;13(1):62.
\newblock doi:{10.1186/s13036-019-0190-3}.

\bibitem{Ghusinga.2017}
Ghusinga KR, Dennehy JJ, Singh A.
\newblock {First-passage time approach to controlling noise in the timing of
  intracellular events}.
\newblock Proceedings of the National Academy of Sciences.
  2017;114(4):693--698.
\newblock doi:{10.1073/pnas.1609012114}.

\bibitem{Timm.2012}
Timm A, Yin J.
\newblock {Kinetics of virus production from single cells}.
\newblock Virology. 2012;424(1):11--17.
\newblock doi:{10.1016/j.virol.2011.12.005}.

\bibitem{Jenner.20181ns}
Jenner A, Yun CO, Yoon A, Kim PS, Coster ACF.
\newblock {Modelling heterogeneity in viral-tumour dynamics: The effects of
  gene-attenuation on viral characteristics}.
\newblock Journal of Theoretical Biology. 2018;454:41--52.
\newblock doi:{10.1016/j.jtbi.2018.05.030}.

\end{thebibliography}

	}

\end{document}